# Efficient Market Dynamics: Unraveling Informational Efficiency in UK Horse Racing Betting Markets Through Betfair's Time Series Analysis

**Narayan Tondapu**
narayan.tondapu@gmail.com
Microsoft, Redmond, Washington, USA

**Abstract**

Using Betfair's time series data, an analysis of the United Kingdom (UK) horse racing market reveals an interesting paradox: a market with short tails, rapidly decaying autocorrelations, and no long-term memory. There seems to be a remarkably high level of informational efficiency in betting exchange returns, in contrast to financial assets that are characterized by heavy tails and volatility clustering. The generalized Gaussian unconditional distribution with a light tail points to a market where knowledge is quickly assimilated and reflected in prices. This is further supported by the extremely quick fading of autocorrelations and the absence of gain-loss asymmetry. Therefore, in addition to measuring long-range memory, the Hurst exponent also shows mean reversion, a sign that markets respond quickly to fresh information.

## 1. Introduction

Economics and finance are based on "stylized facts"—simplified truths extracted from data that may reveal general patterns but also obscure subtleties [1]–[4]. It is arguable that "they are stylized, though whether facts", as Solow quips. However, with careful derivation, these facts become indispensable tools for developing and verifying economic models. The gaming business is being shaken by a 21st-century revolution: internet betting exchanges [5]. Millions of people have placed bets "for" or "against" one another on peer-to-peer platforms like Betfair[1] and Smarkets[2], which have supplanted the conventional bookmaker. The data structure combines and anonymizes bets, emulating financial exchanges like a Limit Order Book (LOB). Examining the domain of online sports betting exchanges reveals a wealth of uncharted terrain about financial market behaviour. Financial markets have been studied extensively for many years, producing a well-established body of "stylized facts", yet there is comparatively less research on online exchanges. The growing popularity of these platforms, their remarkable similarities to traditional financial markets, and this knowledge gap make for an intriguing research opportunity. To begin with, comprehending the stylized facts of betting exchanges provides a special perspective on the financial markets themselves. Therefore, via this study, we can learn new things about the underlying dynamics that influence market behaviour by contrasting and comparing these emerging markets with their more established counterparts. This exchange of ideas has the potential to improve the comprehension of the betting and financial markets, generating more reliable models and perhaps even revealing new trading tactics. Secondly, the process of figuring out the statistical fingerprints of betting exchange data opens the door to the creation of highly advanced artificial data generators. Reliable and authentic data is necessary for these generators, which are essential for training and validating models. Similarly, we can provide these generators with the required statistical legitimacy by obtaining the stylized facts of betting exchanges, which will make them indispensable resources for both practitioners and researchers. Lastly, algorithmic betting will directly benefit from this research. While real betting exchange data is often scarce, costly, and insufficient for some Machine Learning (ML) techniques [6]–[15], large datasets are necessary for training and optimising these algorithms. This research gives players and betting organisations access to realistic synthetic data, which they can use to refine and develop their algorithmic methods. This could lead to more equitable betting practices and a more dynamic betting community.

The following is the paper: We will see the background of our study in the following section. The works that are connected to our method are covered in Section 3. Preprocessing is used to analyze the data in Section 4. The experimental analysis is carried out in Section 5. We include a discussion section in Section 6, and we wrap up the paper with some conclusions and ideas for future research in Section 7.

---

[1] https://www.betfair.com/exchange/plus/
[2] https://smarkets.com/

## 2. Background

### 2.1 Online Sports Betting Exchange

Within the quickly expanding online betting market, betting exchanges have established a distinct market by providing a thriving and maybe more lucrative option than traditional bookmakers. By letting users place bets and lay wagers on a range of event outcomes, these platforms function as markets where users can set their own odds and communicate with one another directly. There are several significant benefits for users of this peer-to-peer approach over the traditional bookmaker model, which consists of setting fixed odds and accepting bets directly from clients. First off, compared to traditional bookies, betting exchanges provide much better odds. The reason for this is that they only receive money as a fee on winning bets, not from the risk of taking wagers. For profitable gamblers, this means increased profits, which is why experienced players and those looking to get the most out of their wagers find the exchange model appealing. Second, betting exchanges help with in-play betting, which enables players to place bets at any time during an event, not just before it begins. Because participants can modify their bets in response to the event as it unfolds, this dynamic element offers an additional layer of excitement and strategic possibility. A more engaging and responsive betting experience is enhanced by real-time odds updates that are depending on the status of the event. Thirdly, a more information-efficient market is encouraged by the exchange concept. Betting exchanges quickly determine which representation of the event's probability distribution is most accurate by combining the knowledge of many different bettors. Both players, who can obtain more dependable odds, and established bookmakers, who can more successfully hedge their bets using the exchange data, gain from this price discovery method. Nevertheless, betting exchanges haven't entirely supplanted conventional bookmakers despite their benefits. Many users still find comfort in the familiarity and simplicity of the bookmaker model, particularly those who are less adept at navigating the exchange interface and setting their own odds. Furthermore, individuals who are risk averse may become discouraged by the exchange model's increased volatility and unpredictability. Even with these drawbacks, the emergence of betting exchanges marks a dramatic transformation in the online gambling market. They provide a strong alternative to the conventional model by emphasising competitive odds, dynamic in-play betting, and information efficiency. This draws in experienced players and fosters a more open and possibly equitable betting environment. Betting exchanges are expected to become ever more crucial as the business develops, influencing how we place bets on sports and other outcomes in the future.

### 2.2 Similarities between Betting Exchanges and Other Market Types

Betting exchanges provide an intriguing window into a world that resembles the busy bustle of financial markets with their dynamic platforms and peer-to-peer trade. These parallels are deeper than superficial ones; they are rooted in the fundamental design and workings of both systems. Fundamentally, betting exchanges and financial markets are just marketplaces where assets can be exchanged. These assets are known as stocks, bonds, and other financial instruments in the world of finance. The "assets" in the context of betting exchanges are wagers on the results of events, mainly athletic ones. The way both platforms enable communication between buyers and sellers highlights this essential similarity. A crucial component that unites both is the existence of a central platform that serves as a middleman for parties engaged in different forms of trade. The current supply and demand for the asset in issue are shown on this platform, which is referred to as the LOB in financial markets and just the "market" in betting exchanges. Buyers enter the LOB by submitting bid orders, which include the requested quantity and the amount they are ready to pay for an item. On the other hand, ask orders are submitted by sellers, who indicate the amount available and the price they are willing to sell for. After that, the matching engine matches vendors and consumers according to their orders, making sure the best deals are carried out. However, there is a slight variation in the way betting exchanges function. Users put bets by "backing" or "laying" an outcome at particular odds and stakes, as opposed to using bid and ask pricing. A winning layer keeps the stake of the lost backer, and the exchange's matching engine matches these bets according to odds and stake, guaranteeing that a winning backer gets their stake multiplied by the winning odds. The basic data design of both markets are further comparable. The way bets are shown on betting exchange markets is quite similar to the LOB structure, which consists of an ordered list of buy and sell orders. Both offer a real-time mood of the market, enabling traders to decide wisely depending on the state of supply and demand. Although there are unquestionable parallels between betting exchanges and financial markets, there are also some significant distinctions that should be noted. Betting exchanges mostly concentrate on making predictions about the course of events, in contrast to financial markets, where the main objective is to make money through dividends or asset appreciation. Variations in forecasting efficiency are caused by this purpose disparity. Although betting exchanges can offer insightful information about the likelihood of an event, they might not be as effective as prediction markets that primarily use binary option trading to produce precise forecasts. By identifying these similarities, we can advance our knowledge of the betting and financial markets, encourage creative thinking in the field, and perhaps even bridge the gap between these two mutually exclusive topics.

### 2.3 Stylized Facts in Financial Time Series Data

Researchers have always been fascinated by financial markets because of their deep dynamics and diverse ecosystems. Discovering and analysing "stylized facts"—repeated statistical patterns that surface from financial time series data—has proven to be one of the most profitable research directions. These subtle indicators of market behaviour provide important new understandings of the fundamental factors influencing price changes and investor choices. Financial data is distinguished by its different nature. Transactions happen at random intervals, and prices fluctuate in modest amounts, usually expressed in cents for stocks. For modelling and analysis, this non-uniformity poses a special difficulty. Autocorrelation is an important term in the study of market dynamics. This quantifies the relationship between a variable and itself at a later period. Surprisingly, financial returns show minor adjustments, suggesting that the market is not very memory-rich. According to the efficient market hypothesis, historical price fluctuations are not very indicative of future changes. But whether we look at the absolute value or squared values of returns, the picture gets more complex. Known as nonlinear autocorrelation, these exhibit a significant positive autocorrelation with a gradual decrease. This means that significant price fluctuations—regardless of their direction—usually follow other significant moves, indicating a clustering of volatility across time. The volatility clustering itself is a notable additional feature. This speaks to the propensity for successively high-volatility periods to occur after one another and vice versa. The "shock persistence" is accounted for by conditional heteroskedasticity (the variance of returns varies with time) in models such as GARCH and ARCH. It is also important to consider the link between returns and predicted volatility. Significant losses frequently result in a rise in volatility in the future due to a negative connection known as the leverage effect. Negative returns raise risk and consequently volatility, which is how financial leverage operates.

Time series data related to finance frequently have non-stationary characteristics, indicating that their statistical properties fluctuate with time [16]–[20]. This non-stationarity causes variations in the standard deviation of returns due to trends, seasonalities, and volatility clustering. On the other hand, it is commonly believed that returns are weakly stationary, which means that their mean and auto-covariance functions stay unchanged. Seasonality is yet another fascinating aspect of financial information. Intraday and intraweek volatility often form U-shaped patterns, peaking in proximity to the open and close of the market. There are also weekday impacts, when particular days have larger average returns. A normal distribution is much different from the unconditional distribution of financial returns. In contrast, it exhibits heavy tails, indicating that the likelihood of severe events exceeds that of a normal distribution. The tail index measures this and for financial time series it exhibits a power-law degradation. It's interesting to note that as populations grow, their heavy-tailed tendencies tend to decrease. Lastly, gain-loss asymmetry characterises the propensity for positive price changes to occur more frequently but at a lesser magnitude, whereas negative price changes occur less frequently but at a bigger scale. This fact implies that traders are more sensitive to losses than gains, which is consistent with ideas like loss aversion. It is essential for anyone attempting to navigate the complicated world of financial markets to comprehend these stylized facts. They provide insightful information about risk assessment, market behaviour, and possible trading tactics. We can better comprehend the dynamic forces that build financial landscapes by deciphering these statistical fingerprints, which will ultimately enable us to participate in this dynamic ecosystem with greater knowledge and authority.

## 3. Related Works

Researchers in a variety of disciplines, including computer science, finance, and economics, have become interested in online betting exchanges due to their quick rise in popularity. So, since they first appeared in the early 2000s, online sports betting exchanges have been examined from a variety of aspects which will be discussed in the following subsections.

### 3.1 A Brief Overview of Various Works on Online Sports Betting Exchanges

For instance, a great deal of research has been done on the behavior of human bettors. It appears that a significant amount of this field's study focuses on problem gambling and its effects. Comprising research works by [21]–[23], and numerous more. These studies frequently seek to identify the features of the gambling platforms that drive and impact the behavior of problem gamblers, in addition to the characteristics of their behavior on online betting exchanges. One work from 2022 that provides an example of this is [24], in which they qualitatively evaluate the effects of developments in online betting platforms during the previous ten years on the behaviors of certain gamblers. As [25] noted, there are a lot of empirical research on the behavior of human bettors that make use of prospect theory and the idea of a representative bettor. The findings of other investigations are characterized as stylized truths regarding the behavior of human bettors. [26] discovered that in a 2022 experiment with frequent sports bettors, respondents were more likely to overestimate their chances of winning when the sports framing was used. However, fixed-odds bets with a bookmaker were employed in the trial instead of an internet betting market, and it was limited to low-income players. Using mathematical or algorithmic techniques to make money trading or betting on betting markets appears to be another well-liked area of study. [27], [28] are a few examples. Lastly, a number of studies have been written about the place of sports betting platforms in the economy and their integration with the gaming, sports, and digital industries. Examples include the SWOT analysis of online betting exchanges conducted by [29] and the analysis of the rivalry and relationship between Betfair and traditional bookmakers by [30].

**3.2 Stylized Facts of Online Sports Betting Data**

The field of academic research on online sports betting exchanges is far from unexplored; however, to the best of our knowledge, very little work has been done on the topic of stylized facts in online sports betting exchanges, comparable to the extensive body of work on stylized facts in financial time series data. Since the 1960s, the stylized realities of financial markets have been investigated; in 1963, [31] identified strong tails in the asset return distribution. Even while internet sports betting markets are relatively new, there is still a significant information gap.

**3.2.1 Favorite Longshot Bias**

Nevertheless, one specific aspect of online sports betting exchanges—the existence and kind of the favorite-lonsghot bias—has been thoroughly studied and well-established. The preferred longshot bias is a well-researched phenomenon that exists in betting and financial markets. It can be defined as gamblers underestimating the favorites, or competitors with the best chance of winning, and overvaluing the longshots, or outsiders and contenders who are thought to be unlikely to win. Although there are fewer studies on this bias in sports betting exchanges than in financial markets such as [32], and [33] paper on the favorite-longshot bias in UK football markets are just a few examples. The majority of them concur that sports betting data exhibits a particularly strong favorite-longshot bias. Its causes, however, are still up for discussion. Some academics see it as an indication of market inefficiency, while others believe it may be due to insider knowledge that certain traders may have.

**3.2.2 Informational Efficiency**

More generally, some studies looks at the informational efficiency of sports betting markets. [34] find that betting exchanges are more informationally efficient than dealer markets and, on average, offer much more competitive pricing using football data from Ladbrokes, William Hill, and Betfair. According to [35], betting exchanges provide better forecast accuracy than alternative market structures. Numerous other studies, such as the previously cited 2007 work by [32], have produced identical results. [36] looks at whether sports betting exchanges are semi-strong form efficient, or if new information is quickly and fully absorbed into betting pricing. They come to the conclusion that they are not, at least not for now. The study's data comes from a football market with indications of bias in favor of the home team's winning chances. Using data from UK horse racing markets, [37] investigate the favourite longshot bias as well as market efficiency more generally. They come to the conclusion that exchanges display both poor and strong form market efficiency, in contrast to conventional bookies and financial markets.

**3.2.3 Other Stylized Facts**

The search for comparable insights in sports betting exchanges is still completely unexplored, despite the wide panorama of stylized facts seen in financial markets. The research on football markets by [38] is a noteworthy exception, as it reveals short-memory processes in trading volumes but long-range correlations in return magnitudes. Moreover, their research employing Detrended Fluctuation reveals self-affine mean-reversion of indicated probabilities. [39] thesis is noteworthy for providing a wider range of stylized facts [7]. After analysing more than 12,000 horse races on Betfair, he discovers anomalous, extremely leptokurtic price increases, growing odds dispersion during competitions, and erroneous horse ranking by wagerers in sizable fields. Unexpectedly, [39] uses Gini, Theil, and Generalised entropy indices to investigate uncertainty evolution, confirming the favorite-longshot bias and pointing to an econophysics method that differs from traditional economic time series research. These innovative studies demonstrate how sports betting data may be used to find rich statistical fingerprints, opening the door to a greater comprehension of these dynamic markets.

**3.3 Stylized Facts of Prediction Market Data**

Even while financial markets are closely related to sports betting exchanges, they cannot provide a comprehensive guide to comprehending their statistical behaviour. [40], which examined online political prediction markets, is a good illustration of this. [40] discovered a rightward tilt in prediction market returns, in contrast to the left-skewed return distributions that characterise financial time series. This one study illustrates how different statistical patterns may be present in sports betting exchanges. But one important distinction sets them apart: binary results. Predictive sites such as PredictIt deal in Arrow-Debreu securities and pay out either $1 or $0 depending on the outcome of an event, in contrast to conventional financial markets. This means, as [40] correctly notes, that returns must be examined directly, as opposed to the more popular approaches of employing logarithmic or absolute returns in exchanges for sports betting and financial markets. This being said, [40]'s work is still a vital source of inspiration for this study. We're exploring an exchange similar to his that has little statistical investigation but is reminiscent of the financial markets. A deeper comprehension of the distinct statistical fingerprints concealed inside the dynamic realm of sports betting exchanges is made possible by the change from the norm.

## 4. Data

This study makes use of an extensive collection of horse racing data from Betfair. The data, which includes 1,056,766 price change signals, 73 marketplaces, and 10 events, provides high-resolution insights with messages being transmitted every 50 milliseconds. The dynamic character of betting activity was highlighted by the average of 9.86 runners (range from 3 to 21) with new bets being matched approximately every 50 seconds (with a standard deviation of 450 seconds) in each market. This extensive dataset offers a strong starting point for investigating the statistical features of online sports betting markets.

### 4.1 Initial Data Format

Betfair provided the data that was used for the analysis. The betting platform provides comprehensive data from the exchange that is *timestamped*. The three package plans are called PRO, BASIC, and ADVANCED, and they are priced differently. They vary in terms of both content and frequency. The PRO plan, which offers the most extensive range of content at the maximum frequency—that is, tick-by-tick—is the source of the data used in this study. Users can download in the *tar.bz2* files containing the Betfair historical data files. The directory has data from a month's worth of trade, and directories categorized by event can be found inside the main folder. Each of the *.bz2* files found in the event directories corresponds to a certain market for that event. Market data in JSON format can be found in the market files. The files actually contain a list of market change messages that list each and every change in the market. *MarketDefinition* and *RunnerChange* messages are the two categories of communications that were captured. The former provide the specifics of the market and record any modifications, including the quantity of active runners, whether the market is in play at the moment, and its status. The latter explain modifications to a runner's specifics, including costs. A fresh *RunnerChange* message is issued each time the price of any runner changes. In Tables I, II, and III, the message format is described in full.

TABLE I
THE ANNOUNCEMENT OF A CHANGE IN THE MARKET. EVERY MESSAGE HAS THESE FIELDS, WHICH ARE FOLLOWED BY A RUNNER CHANGE MESSAGE (TABLE III) OR A MARKET DEFINITION (TABLE II)

| Abbreviation | Description |
|---|---|
| op | Operation type |
| clk | Sequence token |
| pt | Published Time - in milliseconds since the start of the epoch |
| mc | The market change follows this field |
| id | The market's unique identifier |

TABLE II
MARKET DEFINITION FIELDS

| Abbreviation | Description |
|---|---|
| Id | The unique identifier of the market |
| Venue | The name of the venue the event is held at |
| bspMarket | Whether the market supports Betfair SP betting or not |
| turnInPlayEnabled | Whether the market is set to turn in-play or not |
| persistenceEnabled | Whether the market supports 'Keep' bets if the market is to be turned in-play |
| marketBaseRate | The commission rate applicable to the market |
| eventId | The unique identifier for the event |
| eventTypeId | The unique eventTypeId that the event belongs to |
| numberOfWinners | The number of winners in the market |
| bettingType | The betting type of the market. One of ODDS, ASIAN HANDICAP DOUBLE LINE or ASIAN HANDICAP SINGLE LINE |
| marketType | The market base type |
| marketTime | The market start time |
| suspendTime | The market suspend time |
| bspReconciled | Whether the market starting price has been reconciled |
| complete | Whether runners can still be added to the market or not |
| inPlay | Whether the market is currently in play |
| crossMatching | True if cross matching is enabled |
| runnersVoidable | True if runners in the market can be voided |
| numberOfActiveRunners | The number of currently active runners |
| betDelay | The number of seconds an order is held until it is submitted into the market. Orders are usually delayed when the market is in-play |
| status | Market's status, for example OPEN or SUSPENDED |
| regulators | Market's regulators |

| Abbreviation | Description |
|---|---|
| discountAllowed | Whether the users' discount rate is valid on the market or not |
| timezone | The timezone the event is taking place in |
| openDate | The start and end dates of the event. By default GMT |
| version | A number indicating market changes |
| name | Market's name |
| eventName | The name of the event |

TABLE III
RUNNER CHANGE FIELDS

| Abbreviation | Description |
|---|---|
| status | The status of the selection, for example ACIVE or LOSER |
| sortPriority | Runner's sort priority |
| bsp | Runner's Betfair Starting Price |
| removalDate | Date and Time of runner's removal |
| id | The unique selctionId of the runner |
| name | Runner's name |
| hc | Runner's handicap |
| adjustmentFactor | The adjustment factor applied upon selection's removal |
| tv | The total amount matched across the market, the Traded Volume |
| ltp | Last Traded Price |
| spb | Starting Price Back |
| trd | Traded PriceVol |
| spf | Starting Price Far |
| atb | Available to Back |
| spl | Starting price Lay |
| spn | Starting price Near |
| atl | Available to Lay |

### 4.2 Data Preprocessing

This study used a simplified method to retrieve the used high-frequency data straight from JSON files rather than engaging with Betfair's API directly. For the manageable data size of less than 600MB, this proved to be more practical and efficient. The information was converted into Pandas data frames to facilitate handling, and subsequently sifted and processed to produce pertinent CSV files. Because of its smaller data size and emphasis on readability, the well-known and intelligible CSV format was selected over possibly quicker binary formats like pickle or feather. For flexibility, runner change messages were separated further by event and combined into a single file. Messages pertaining to market definition were kept apart; one file contained all modifications, while another offered a succinct view of one message per market. From the market definitions, a second CSV was created with the winner and runner count information. Ultimately, individual CSV files containing the raw, *timestamped* data were obtained from each market and event. This effective data organisation made it easier to do more research and analysis.

#### 4.2.1 Runner Changes

By looking for runner changes (*rc*) in each and every market change message, the runner change data is found. Because of the way the JSON files are formatted, anything that comes after the *mc* field—whether it's a market definition or runner change—loads into the data frame as a single string that must be parsed in order to extract the relevant fields. We then store the result as a string representing GMT time in the format Year-Month-Day (Hour: Minute: Second) after converting the *pt* field to standard Unix epoch time in seconds. Based on their timestamps and market identifiers, the information from the market definition messages was appended to the runner change messages to include the *inPlay* field, which indicates if the market is in-play at that moment. Next, the data frame is arranged in ascending order by time. Then, we are left with the final data format that is displayed in Table IV after removing the fields that we are not interested in. It is important to keep in mind that the majority of the fields in market change messages are nullable and delta based, which means that if they are not modified, they will be set to null.

TABLE IV
THE FINAL STRUCTURE OF THE RUNNER CHANGE DATA

| Field | Description | Data Type |
|---|---|---|
| atb | Available to Back | string |
| id | Selection identifier | int |
| t | Time | string |
| inPlay | Whether the market is currently in-play | bool |
| spn | Starting Price Near | float |
| spf | Starting Price Far | float |
| atl | Available to Lay | string |
| spl | Starting Price Lay | string |
| trd | Traded PriceVol | float |
| ltp | Last Traded Price | float |
| tv | Traded Volume | float |
| spb | Starting Price Back | string |
| eventId | Unique event identifier | int |
| marketId | Unique market identifier | float |

### 4.2.2 Market Definitions

For the purpose of clarity and conciseness, the market definition messages have been effectively divided into two distinct files. All changes to the market definition over the course of their lifetimes are scrupulously tracked in one file, offering a comprehensive historical record. The other file takes a more condensed approach, giving each market a single representative definition. Notably, the "*inPlay*" field is the only difference between the two formats' identical structure, as is seen in Table V. Only available in the whole change file, this boolean field specifies whether the market is open for business at this time during the event.

TABLE V
THE FINAL STRUCTURE OF THE MARKET DEFINITION DATA

| Field | Description | Data Type |
|---|---|---|
| id | The unique identifier of the market | float |
| turnInPlayEnabled | Whether the market is set to turn in-play or not | bool |
| marketBaseRate | The commission rate applicable to the market | float |
| eventId | Unique event identifier | int |
| marketTime | Time | string |
| suspendTime | The date and time the market is due to be suspended on | string |
| complete | Whether runners can still be added to the market or not | bool |
| numberOfActiveRunners | Number of runners currently available to bet on | int |

### 4.2.3 Winners

Extracted from the market definition messages, a dedicated file consolidates winner information for each market. This file originated from runner statuses like "ACTIVE", "REMOVED", "WINNER", "PLACED", "LOSER", or "HIDDEN" within the market definitions. The structure of this winner file is outlined in Table VI for the reference.

TABLE VI
THE FINAL STRUCTURE OF THE WINNERS DATA

| Field | Description | Data Type |
|---|---|---|
| id | The unique identifier of the market | float |
| winner | The identifier of the winning runner | int |
| eventId | Unique event identifier | int |
| numberOfRunners | Total number of runners that was available to bet on while the market was open | int |

### 4.2.4 Returns

For future distributional comparisons, positive and negative data were segregated in order to compute and export *timestamped* returns. The final runner status, traded volume, and last traded price were used to calculate returns. For the purpose of capturing both good and negative outcomes, each price signal included returns for both the "back" and "lay" sides. We took into account commission fees, accounting for both the standard 5% rate and variances based on winnings. Due to commission deductions, the computations took into account win/loss scenarios for both parties and produced net returns. Changes in trade volume for particular runner picks were analysed to determine the stake. With respect to positive and negative returns, distinct files were exported for every event. While keeping timestamps, these might then be concatenated into a combined returns file. For evaluations that were particular to a given market, markets IDs were included. A compromise was found between preventing unnecessary file loading during analysis and providing easily navigable data chunks for preliminary exploration while storing data by event. The transformation of returns was done in a logarithmic, simple, squared, or raw or absolute form, depending on the particular properties being investigated and the methodologies selected. During data loading and processing, these transformations were used when necessary.

## 5. Experimental Analysis

This section will outline the approach and procedure used to extract stylized facts from data from online sports betting exchanges. Along with that, we'll demonstrate and talk about the outcomes.

### 5.1 The Distributed Properties of Returns

In this section, we look into the basic properties of log returns in our betting data. We examine whether "heavy tails", an indicator of severe occurrences frequently observed in financial markets, are present or absent as we close in on the distribution of these returns. We may comprehend the form and behaviour of the distribution by figuring out the tail exponent and fitting suitable functions. To further investigate biases in the distribution of wins and losses inside the betting exchange, we also look at gain-loss asymmetry by examining the distributions of positive and negative returns independently. Important insights into the underlying mechanisms of returns and any potential departures from well-established financial market trends will be gained from this study. Generally, logarithmic returns can be defined as according to Equation (1).

$$R_t = ln(P_t) - ln(P_t - 1) \tag{1}$$

where $R_t$ denotes the logarithmic returns from a period of time $t$, $P_t$ stand for the price from the period $t$, and $P_t - 1$ stands for the price from the previous period, $t - 1$. Log returns provide an approximation for simple returns and the relationship between the two is shown in Equations (2) and (3).

$$R_t = ln(1 + r_t) \tag{2}$$

$$r_t = exp(R_t) - 1 \tag{3}$$

where $R_t$ denotes log returns and $r_t$ stand for simple returns. Generally, simple returns are defined as according Equation (4).

$$r_t = P_t - P_{t-1}/P_{t-1} \tag{4}$$

Log returns offer numerous significant advantages for time series analysis in this context, despite the fact that basic or raw returns may seem obvious. First of all, unlike simple returns that aggregate across assets and may cause misunderstandings, log returns provide meaningful addition over time periods. This is more consistent with the multiplicative character of price changes over time. Second, since log scales keep minor values from becoming lost at the bottom of the graph, visualising financial data is made simpler and more understandable with log returns. Thirdly, unlike simple returns, log returns exhibit symmetry around zero, which means that positive and negative changes of the same size cancel each other out. Lastly, a primary goal of this study is to compare sports betting exchanges to regular markets using "stylized facts" from established financial markets, which are mostly generated from log returns. We make use of the integrated data from all marketplaces and events to guarantee reliable analysis. This method minimises potential biases from individual market segmentation while maximising the strength of the data. We can learn a great deal about the dynamics and features of returns in the betting exchange ecosystem by examining the distribution of log returns.

#### 5.1.1 The Unconditional Distributions of Returns

We examine the return distribution across 41588 time intervals. Table VII illustrates that the distribution's mean is in close proximity to zero, suggesting that the overall volume of both positive and negative returns is almost equal. The gain-loss asymmetry must still be investigated in next section since their distributions may still differ. Given the standard deviation of 4.0323, the variation coefficient $c_v = \mu/\sigma$ equals $c_v = -0.0004$, a value that is substantially smaller than the one empirically found, for instance, by [41], in financial market data. Additionally, [41] discovered negative skewness, which the Betfair data does not exhibit. Despite having the opposite sign, the skewness is comparable in size. Predictive market data likewise showed positive skewness, but much higher. This suggests a gain-loss asymmetry that differs from what is typically observed in data from financial markets.

TABLE VII
THE UNCONDITIONAL DISTRIBUTION OF LOG RETURNS' DESCRIPTIVE STATISTICS

| Number of Observations | Mean | Standard Deviation. | Skewness | Kurtosis |
|---|---|---|---|---|
| 41588 | -0.0018 | 4.0323 | 0.0241 | 1.0994 |

#### 5.1.2 Tails

The finding with the low kurtosis is a little unexpected. According to Pearson's definition, a normal distribution has a kurtosis of 3.0. This is what we use. This indicates that, in comparison to a normal distribution, the data is platykurtic, with a wider peak and fatter tails. The outcome deviates from previous understanding of financial timeseries that has had a strong tails distribution in the unconditional distribution of financial asset returns for more than 50 years. [40] has also discovered evidence of big tails in prediction markets. In order to examine the tail distribution in more detail, we compute the tail-index, or $\alpha$, which is a measure of the tail's weight. The tail gets thinner as $\alpha$ increases. The Hill estimator [42] is one technique for calculating the tail-index of a distribution's tails. It is an estimate of quasi-maximum likelihood. The number of tail observations ($k$) employed has a significant impact on the estimator's performance. Depending on the value of $k$, the estimator's bias and variance are subject to trade-offs. In particular, if the distribution has a smaller sample size or the tail behavior is uncertain, a smaller number lowers the risk of bias and overfitting. However, if the distribution has heavy tails, a greater value of $k$ usually yields a more accurate estimate of the tail index. Additionally, it can strengthen the estimator's resilience and lower its variability. The selection of $k$ is a subject of continuous discussion. While "eye-balling" the right number of tail observations is not unusual, selecting the $k$ that minimizes the Mean Squared Error (MSE) is the most widely used method. Instead, [43] suggest a strategy centered on fitting the tail by reducing the quantile dimension's maximum deviation. Additional estimators consist of the Ratio estimator and Pickand's tail-index estimator. As a percentage of the total number of observations in the distribution, $k$ is represented in Table IX. Table IX displays the Hill estimator values for the Betfair data. A tail-index of three is closely correlated with a hill estimator of value -0.3. However, the approach works best with heavy-tailed distributions, therefore the result may not be very accurate for our data. A generalized normal distribution with a $\beta$ value of 1.19 was found to have the minimum sum squared error when several distributions were fitted to the data as shown in Fig. 1. When $\beta = 1$, the distribution resembles a Laplace distribution; when $\beta = 2$, it resembles a Gaussian distribution. For a generalized Gaussian, the probability density function is provided by according to Equation (5).

$$f(x, \beta) = \beta/2T\left(\frac{1}{\beta}\right) \exp\left(-|x^\beta|\right) \tag{5}$$

TABLE IX
THE VALUES OF THE HILL ESTIMATOR

| $k$ | Hill estimator |
|---|---|
| 0.01 | -0.348 |
| 0.02 | -0.454 |
| 0.03 | -0.533 |
| 0.04 | -0.599 |
| 0.05 | -0.656 |
| 0.1 | -0.877 |

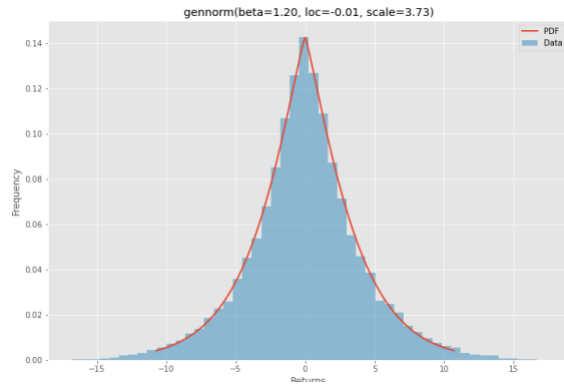

Fig. 1. A generalized normal distribution fitted to the data's probability density function

### 5.1.3 Gain Loss Asymmetry

We compare the unconditional distributions of the positive and negative returns to look at the symmetry between them. The distributions' descriptive statistics are displayed in Table X. The numbers are extremely close, and the two distributions, as shown in IX, are almost identical when viewed visually. We do a two-sample Kolmogorov-Smirnov (KS) test in order to examine more. The two samples are taken from the same underlying distribution, which is the null hypothesis $H_0$. The greatest vertical distance between the two empirical distribution functions is the KS test statistic $D$. If the $p$-value is sufficiently small or the value of the statistic $D$ is greater than the critical value $D_c$, which is defined as according to Equation (6).

$$D_c = c(\alpha)\sqrt{\frac{n_a+n_b}{n_a n_b}} \tag{6}$$

where $n_a$ and $n_b$ are the numbers of observations in the distributions that are being compared, and $c(\alpha)$ is a constant that represents the inverse of the Kolmogorov distribution at a significance level $\alpha$ (often taken to be 0.05). We obtain $D_c = 0.0080$ for $\alpha = 0.05$, $D =$

0.0068, and $p$ value = 0.1347 by computing $D$, $D_c$, and the $p$-value. We are unable to rule out the null hypothesis that the positive and negative returns have the same distribution since $D < D_c$ and $p > 0.05$. This would suggest that there is no gain-loss asymmetry because the gains and losses are roughly equal in size. Generalized normal distributions with values of $\beta$ near to those in the distribution of total returns were the distributions that produced the fewest mean squared errors, both for positive and negative returns. It is depicted in Fig. 2.

TABLE X
THE UNCONDITIONAL DISTRIBUTIONS OF POSITIVE AND NEGATIVE RETURNS AND THEIR DESCRIPTIVE STATISTICS

| Returns | Number of Observations | Mean | Standard Deviation | Skewness | Kurtosis |
|---|---|---|---|---|---|
| Positive | 57648 | 0.0001 | 3.769 | 0.0036 | 1.362 |
| Negative | 57648 | 0.0001 | 3.628 | 0.0102 | 1.325 |

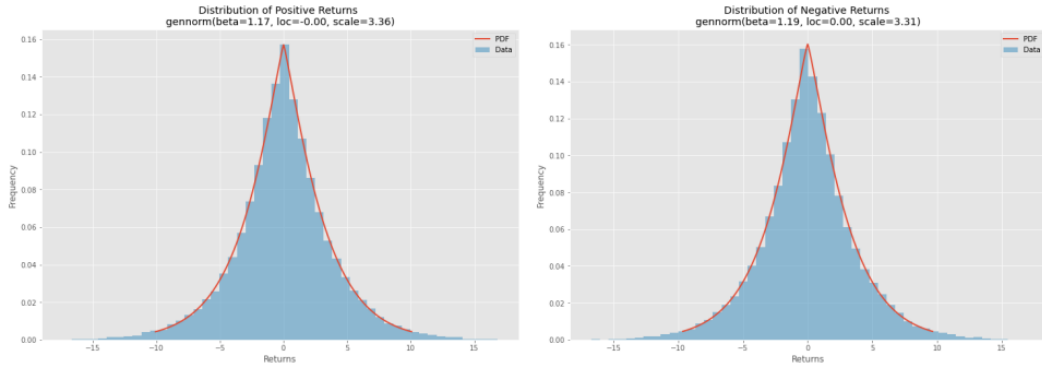

Fig. 2. A fitting of the probability density function of positive and negative returns using a generalized normal distribution

## 5.2 The Time Dependence Properties of Returns

Researchers have been particularly interested in the time dependence features of financial time series because they add to the conversation about the market efficiency of financial exchanges. In contrast to the distributional features of the data, we use analysis of absolute and squared returns in addition to log returns. Furthermore, rather than examining the entire dataset at once, we compute and report the results for each market independently. This decision was made because investigating the possibility of information spreading across markets, even in the unlikely event that it did, would be outside the purview of this investigation. When determining stylized facts related to temporal dependency, studies on the dependence qualities of financial time series data typically concentrate on a single market at a time. This is because, unlike sports betting markets, the market may operate for multiple years, generating a greater volume of data.

### 5.2.1 Stationarity

Prior to delving into the in-depth examination of time dependency features, we must guarantee a vital prerequisite: stationarity. This basically means that there are no time-dependent patterns in the return process, which gives us the confidence to mix data from different time periods. Formally, under any temporal shift, a stationary signal has an independent joint Cumulative Distribution Function (also known as an independent CDF). Our two tests of choice are the Kwiatkowski-Phillips-Schmidt-Shin (KPSS) (Table XI) and Augmented Dickey-Fuller (ADF) (Table XII) tests. The null hypothesis ($H_0$) states that unit roots are present (non-stationarity), while the alternative ($H_1$) states that they are absent (stationarity). The ADF test is designed to address unit roots. Therefore, before examining time-dependent features, stationarity testing becomes crucial. In order to identify unit roots in time series data while taking serial correlation into consideration, we utilise the ADF test as shown in Table XII. Given that a unit root is present, non-stationarity is implied by the null hypothesis ($H_0$). On the other hand, stationarity is indicated by the alternative hypothesis ($H_1$). A higher test statistic that is negative suggests a more robust rejection of the null hypothesis. The $p$-value must be less than 0.05 and the statistic must drop below a crucial value in order to obtain this rejection. We used three example markets from the Betfair data to apply the ADF test, concentrating on absolute returns (important for further study). For the remaining markets, similar outcomes were seen, with continuously low $p$-values and statistics. This verifies that there are no unit roots in any of the 73 marketplaces and decisively rejects the null hypothesis. This result clearly suggests that the data are trend-stationary, which is a prerequisite for our intended examination of the time dependent features. Significantly rejecting $H_0$, trend-stationarity is confirmed by highly negative test statistics and low $p$-values (found in all 73 markets). In order to achieve strict stationarity, further differencing is necessary even though the data shows a deterministic trend without unit roots. The ADF analysis is enhanced by the KPSS test (Table XI), which has an opposing null hypothesis ($H_0$: weakly stationary). Here, $H_0$ is continuously rejected, demonstrating the stationarity of differences. This indicates that following differencing, the data becomes stationary, supporting the trend-stationarity finding from the ADF test. It's interesting to note that stationarity is shown by running both tests on log returns, as Tables XIII and XIV. This implies that stationarity properties may be introduced by log transformations, which calls for more research in subsequent studies. Using both ADF and KPSS tests, we are able to obtain a thorough grasp of the stationarity of our data. This sets the groundwork for the subsequent parts' trustworthy examination of time-dependent features.

TABLE XI
THE KPSS TEST RESULTS FOR THREE SAMPLE MARKETS

| Market Id | KPSS Statistic | Critical Value (5%) | P-value | Rejected |
|---|---|---|---|---|
| 1.122946937 | 1.21043 | 0.463 | 0.01 | True |
| 1.122946927 | 2.47760 | 0.463 | 0.01 | True |
| 1.122946942 | 0.98900 | 0.463 | 0.01 | True |

TABLE XII
THE ADF TEST RESULTS FOR THREE SAMPLE MARKETS

| Market Id | ADF Statistic | Critical Value (5%) | P-value | Rejected |
|---|---|---|---|---|
| 1.122946937 | -5.021294 | -2.863 | 0.000020 | True |
| 1.122946927 | -8.518810 | -2.862 | 0.000000 | True |
| 1.122946942 | -5.129096 | -2.864 | 0.000012 | True |

TABLE XIII
THE KPSS TEST RESULTS FOR THREE SAMPLE MARKETS FOR LOG RETURNS

| Market Id | KPSS Statistic | Critical Value (5%) | P-value | Rejected |
|---|---|---|---|---|
| 1.122946937 | 0.014013 | 0.463 | 0.01 | True |
| 1.122946927 | 0.053100 | 0.463 | 0.01 | True |
| 1.122946942 | 0.049614 | 0.463 | 0.01 | True |

TABLE XIV
THE ADF TEST RESULTS FOR THREE SAMPLE MARKETS FOR LOG RETURNS

| Market Id | ADF Statistic | Critical Value (5%) | P-value | Rejected |
|---|---|---|---|---|
| 1.122946937 | -13.735484 | -2.863 | 0.00000 | True |
| 1.122946927 | -20.866090 | -2.862 | 0.00000 | True |
| 1.122946942 | -10.775032 | -2.864 | 0.00000 | True |

### 5.2.2 Autocorrelations

We examine the autocorrelations of asset returns to evaluate market efficiency and temporal dependency. A variable's present value's relationship to its historical values is measured by autocorrelation. High-frequency financial data is well recognised for lacking considerable linear autocorrelation, with the exception of brief lags that show a negative trend that is progressively declining and a negative spike at the beginning of the lag. For Betfair data, this trend is consistent with what was previously shown for stationary log returns. Visual examination of autocorrelation plots for tick-by-tick returns for several markets validates this pattern. The plots, which are displayed in Fig. 3, have a large negative autocorrelation at the first lag and may continue to do so for a few more lags until becoming negligible. This suggests market efficiency because there is no discernible autocorrelation. Correlation decay rate is a measure of how quickly the market reacts to fresh information. Market efficiency would result from traders swiftly nullifying exploitable correlations in search of successful methods. The Efficient Market Hypothesis (EMH) has been the subject of decades-long controversy, but despite this, most scholars agree that financial transactions involve some level of efficiency. There appears to be a comparable degree of information efficiency in online sports betting exchanges, as evidenced by the tight correlation between the reported autocorrelation in betting exchange data and that observed in liquid financial markets.

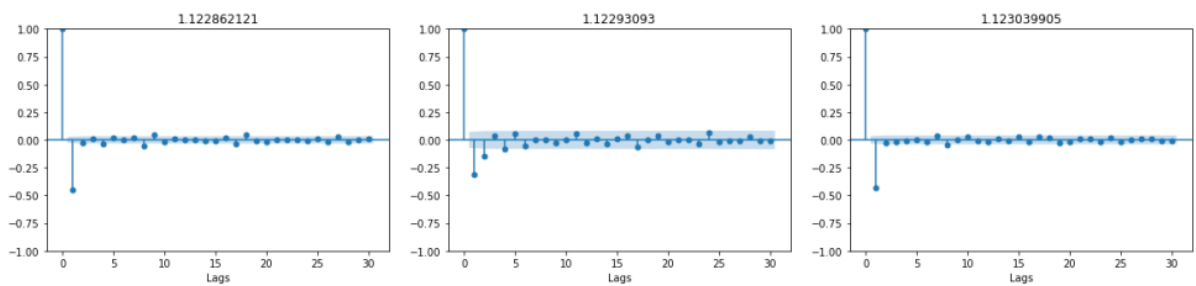

Fig. 3. Plots of autocorrelation for three sample markets

### 5.2.3 Volatility Clustering

Volatility clustering suggests that there may be inefficiencies in the market because relatively small price changes are typically followed by similar movements, and large price changes by similar huge ones. The volatility patterns' predictability points to potential for abuse. We examine the non-linear autocorrelation of squared or absolute returns to further investigate this. Financial time series typically show high positive non-linear autocorrelation, which decays slowly and frequently in the form of a power law with an exponent ($\alpha$) between 0.1 and 0.4 as shown in Fig. 4. It's interesting to see that the power-law behaviour varies even if the Betfair data exhibits strong positive non-linear autocorrelation. Table XV shows that the average exponent $\alpha$ is greater than what is usually found in financial markets. This indicates a quicker rate of autocorrelation decay, which may indicate more market efficiency in the betting exchange.

TABLE XV
THE ABSOLUTE RETURN VALUES OF THE POWER-LAW EXPONENT $\alpha$

| Number of Observations | Mean | Standard Deviation | Max | Min |
| --- | --- | --- | --- | --- |
| 73 | 0.62435 | 0.3052 | 1.9755 | 0.1645 |

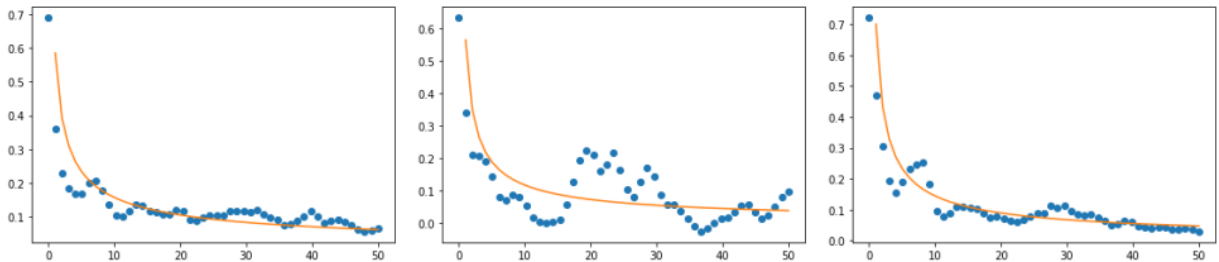

Fig. 4. Plots of autocorrelation for three sample markets fitted with power-law functions

## 6. Discussion

This study explores possible causes and limits as it digs deeper into the study's findings. The return distribution's lack of heavy tails, which stands in sharp contrast to conventional wisdom in the financial markets, calls for more research. One possibility is that athletic events are inherently more predictable than world-changing events that have an impact on financial markets. Lighter-tailed distributions may also be caused by increased market efficiency in betting exchanges, which is indicated by fewer anomalies and a quicker rate of autocorrelation decay. This supports previous studies and offers plausible answers. First, the efficiency of the market may be enhanced by the participation of experienced bettors, who are thought to be less prone to mispricing. Second, compared to financial markets, betting exchanges may draw users who are less risk averse, which could result in a more evenly distributed allocation of profits and losses. However, biases and data limitations lead to limits. There are questions regarding the representativeness of the data due to the very short timeframe (one month). The presence of calendar impacts, which are well-documented in financial markets, must be ruled out using annual or even longer-term data analysis. Furthermore, it is difficult to draw firm conclusions because there isn't much research that directly compares the risk aversion of stock traders and users of betting exchanges. In spite of these drawbacks, the research provides insightful information. The lack of heavy tails implies that there are notable differences between the behaviour of participants in betting exchanges and the dynamics of sporting events when compared to typical financial markets. To fully comprehend these variations and their consequences for market efficiency, participant behaviour, and prospective trading strategies, more research is required using larger datasets and more advanced analysis approaches. We can get a more in-depth understanding of the distinct features and dynamics that differentiate betting exchanges from their financial equivalents by tackling these constraints and conducting further analysis on the data.

## 7. Conclusion and Future Works

Through careful examination of Betfair's horse racing data, we were able to pinpoint a number of crucial traits of online sports betting exchanges. Interestingly, the return distribution showed "light tails", indicating less dramatic events than what is typically seen in financial markets. Furthermore, there was no evidence of gain-loss asymmetry, and the returns showed mean reversion and stationarity. These results suggest that there may be more informational efficiency in betting exchanges, as does the faster decay of autocorrelations. Although the initiative looked at stylized statistics from the existing financial industry, it wasn't comprehensive. Due to time constraints, the leverage effect and aggregational Gaussianity could not be investigated. Furthermore, for a more reliable Hurst exponent estimate, some analyses might benefit from more sophisticated methods like Detrended Fluctuation Analysis. All things considered, this study provides a useful foundation for comprehending the distinct statistical characteristics of betting exchanges, emphasising both their parallels and divergences with financial markets.

Building on the study's first findings, a number of intriguing directions demand more investigation. First off, there are more stylized facts from financial markets that should be looked into in betting exchanges. These include the relationship between volume and volatility, conditional heavy tails, and possible time-scale asymmetries. Second, there is an abundance of opportunity to identify different traits if the study is broadened to include more data, different sports outside horse racing, and in-play versus non-in-play markets. Thirdly, examining seasonal impacts specific to betting exchanges—as opposed to conventional calendar cycles—may provide insightful information. At last, validating the results and opening the door to additional theoretical and practical applications would require reproducing these stylized facts using a synthetic betting exchange data generator such as the Bristol Betting Exchange. Following these paths will help us comprehend the complex dynamics seen in betting exchanges better and will add to our understanding of these intriguing markets.